\documentclass[aps,pre,preprint,a4paper,unsortedaddress,onecolumn]{revtex4}

\usepackage[fleqn]{amsmath}
\usepackage{amssymb}
\usepackage[dvips]{graphicx}
\usepackage{color}
\usepackage{tabularx}
\usepackage{mathtools}
\usepackage{algpseudocode}
\usepackage{enumitem}
\usepackage{epstopdf}


\newcommand{\vect}[1]{\textbf{\textit{#1}}}

\newcommand{\dir}{\mathrm{dir}}
\newcommand{\rec}{\mathrm{rec}}
\newcommand{\corr}{\mathrm{correction}}
\newcommand{\correlation}{\mathrm{corr}}
\newcommand{\erfc}{\mathrm{erfc}}

\newcommand{\ik}{\mathrm{ik}}
\newcommand{\ad}{\mathrm{ad}}
\newcommand{\homo}{\mathrm{homo}}
\newcommand{\inhomo}{\mathrm{inhomo}}

\newcommand{\qh}{q_{\mathrm{H}}}
\newcommand{\qo}{q_{\mathrm{O}}}
\newcommand{\sh}{s_{\mathrm{H}}}
\newcommand{\so}{s_{\mathrm{O}}}

\newcommand{\myphi}{\varphi}

\newcommand{\hmyphi}{\hat{\varphi}}

\newcommand{\wshape}{H}
\newcommand{\mydx}{\Delta x}

\newcommand{\mo}{\mathcal {O}}
\newcommand{\me}{\mathcal {E}}
\newcommand{\mg}{\boldsymbol{\mathcal {G}}}
\newcommand{\mf}{\boldsymbol{\mathcal {F}}}

\newcommand{\imgunit}{\mathrm {i}}

\newcommand {\newparagraph} {\vskip .3cm\noindent}

\begin{document}

\title{The optimal particle-mesh interpolation basis}
\author{Han Wang}
\email{wang_han@iapcm.ac.cn}
\affiliation{Institute of Applied Physics and Computational Mathematics, Fenghao East Road 2, Beijing 100094, P.R.~China}
\affiliation{CAEP Software Center for High Performance Numerical Simulation, Huayuan Road 6, Beijing 100088, P.R.~China}
\author{Jun Fang}
\affiliation{Institute of Applied Physics and Computational Mathematics, Fenghao East Road 2, Beijing 100094, P.R.~China}
\affiliation{CAEP Software Center for High Performance Numerical Simulation, Huayuan Road 6, Beijing 100088, P.R.~China}
\author{Xingyu Gao}
\affiliation{Laboratory of Computational Physics, Huayuan Road 6, Beijing 100088, P.R.~China}
\affiliation{Institute of Applied Physics and Computational Mathematics, Fenghao East Road 2, Beijing 100094, P.R.~China}
\affiliation{CAEP Software Center for High Performance Numerical Simulation, Huayuan Road 6, Beijing 100088, P.R.~China}

\begin{abstract}
  The fast Ewald methods are widely used to compute the point-charge electrostatic interactions in molecular simulations.
  The key step that introduces errors in the computation is the particle-mesh interpolation.
  In this work, the optimal interpolation basis is derived by minimizing the estimated error of the fast Ewald method.
  The basis can be either general or model specific, depending on whether or not the charge correlation has been taken into account.
  By using the TIP3P water as an example system,
  we demonstrate that the general optimal basis is always more accurate than the B-spline basis in the investigated parameter range,
  while the computational cost is at most 5\% more expensive.
  In some cases, the optimal basis is found to be two orders of magnitude more accurate.
  The model specific optimal basis further improves the accuracy of the general optimal basis, but requires more computational effort in the optimization,
  and may not be transferable to systems with different charge correlations. 
    Therefore, the choice between the general and model specific optimal bases is a trade-off between the generality and the accuracy.
\end{abstract}

\maketitle

\section{Introduction}

The computation of the electrostatic interaction is an important and non-trivial task in molecular simulations. 
The difficulty lies in the slow decay of the Coulomb interaction with respect to the particle distance,
thus the cut-off method that ignores the particle interactions beyond a certain range usually leads to
unphysical artifacts~\cite{vanderSpoel2006origin}.
This problem is solved by the Ewald summation~\cite{ewald1921die},
which splits the electrostatic interaction into
a short-ranged particle-particle interaction that is computed by the cut-off method
and an interaction of smeared charges that is computed by solving the Poisson equation in the reciprocal space.
The optimal computational expense of the Ewald summation grows in proportion to the three-halves power of the number of particles,
and is unaffordable for systems that are larger than several hundreds of particles~\cite{pollock1996comments}.

\newparagraph
Instead of the Ewald summation, the fast Ewald methods
are widely used nowadays and implemented in molecular simulation packages~\cite{pronk2013gromacs,phillips2005scalable,plimpton1995fast}.
Some examples are
the smooth particle mesh Ewald (SPME) method~\cite{darden1993pme,essmann1995spm},
the particle-particle particle-mesh (PPPM) method~\cite{hockney1988computer,deserno1998mue1}
and the nonequispaced fast Fourier transform based method~\cite{hedman2006ewald,pippig2013pfft}.
These methods reduce the computational expense to $\mo(N\log N)$ ($N$ being the number of particles) by 
accelerating the solution of the Poisson equation with the fast Fourier transform (FFT).
Although the fast Ewald methods are substantially faster than the Ewald summation, the accuracy is inferior.
The only step that introduces errors
is the particle-mesh interpolation, which interpolates the particle charges on a uniform mesh,
and the solution of the Poisson equation represented on the mesh back to the particles.
Therefore, the quality of the interpolation basis plays an important role in the accuracy of the fast Ewald methods~\cite{deserno1998mue2,wang2010optimizing,ballenegger2012convert,wang2012numerical}.

\newparagraph
All the mentioned fast Ewald methods use the cardinal B-spline basis for the particle-mesh interpolation,
which was proved to be superior to the Lagrangian interpolation basis~\cite{deserno1998mue1}.
Very recently, Nestler~\cite{nestler2016parameter} and Gao et.~al.~\cite{gao2017kaiser}
showed that the Bessel and Kaiser-Bessel bases are more accurate than the B-spline basis
in certain ranges of the working parameter space,
which is spanned by 
the splitting parameter (how the two parts of the Ewald summation are split),
the mesh spacing and the truncation radius of the interpolation basis.
These observations indicate that the B-spline basis that is the ``golden standard''
in the particle-mesh interpolation can be improved, at least, in part of the parameter space.

\newparagraph
In this work, the optimal particle-mesh interpolation basis in the sense of minimizing the estimated error of the fast Ewald method is proposed.
In our approach, the optimal basis is discretized by the cubic Hermite splines,
and the values and derivatives of the basis at the discretization nodes
are adjusted by solving an unconstrained optimization problem.
We prove that, {as long as the system size is large enough}, the optimal interpolation basis {is system independent,
  and  is} determined by a characteristic number defined by the product of the splitting parameter and the mesh spacing.
We numerically investigate the accuracy of the optimal interpolation basis
in a TIP3P water system, 
and demonstrate that the optimal basis always outperforms the B-spline and the Kaiser-Bessel bases in the investigated parameter range.
In some cases, the optimal basis is more than two orders of magnitude more accurate than the B-spline and the Kaiser-Bessel bases.
We also show that the time-to-solution of using the optimal basis is marginally longer
than the B-spline basis, and is no more than the Kaiser-Bessel basis.
We  report that
the accuracy of the optimal basis is further improved by taking into account the
charge correlations during the basis optimization.
However, the derived optimal basis is model specific, and cannot be transferred to the simulation with the same product of the splitting parameter and the mesh spacing.
This implies that {when simulating systems with different charge correlations,
the basis should be re-optimized.
Therefore,
whether or not to consider the charge correlation in the basis optimization is a trade-off between the generality and the accuracy.

\newparagraph
The manuscript is organized as follows:
In Sec.~\ref{sec:ewald}, the fast Ewald method is introduced briefly.
The optimal interpolation basis is proposed in Sec.~\ref{sec:opti}, and the generality of the optimal basis is discussed in Sec.~\ref{sec:generality}.
In Sec.~\ref{sec:result}, the accuracy of the optimal basis is investigated in the TIP3P water system as an example,
and the advantage over the B-spline and the Kaiser-Bessel bases is demonstrated.
In Sec.~\ref{sec:corr}, we show that the accuracy of the optimal basis is further improved, at the cost of generality,
by taking into account the charge correlation in the system.
The work is concluded in Sec.~\ref{sec:conclusion}.

\section{The fast Ewald methods}
\label{sec:ewald}

We consider $N$ point charges that are denoted by $\{q_1, \cdots, q_N\}$ in
a unit cell with periodic boundary condition.
The positions of the charges are denoted by $\{\vect r_1, \cdots, \vect r_N\}$, respectively.
The Coulomb interaction of the unit cell is given by
\begin{align}
  E = \frac12 \sum_{\vect n} ^\ast\sum_{i,j=1}^N \frac{q_i q_j}{\vert \vect r_{ij} + \vect n\vert},
\end{align}
where $\vect r_{ij} = \vect r_i - \vect r_j$, and
$\vect r_{ij} + \vect n$ is the distance between charge $i$ and all periodic images of
charge $j$, because we have
$\vect n = n_1 \vect a_1 + n_2 \vect a_2 + n_3 \vect a_3$
with $(n_1, n_2, n_3) \in \mathbb Z^3$ and $(\vect a_1, \vect a_2, \vect a_3)$
being the unit cell vectors.
The ``$\ast$'' over the outer summation means that when $\vect n=0$, the $i=j$ terms should be skipped in the inner summation.
The prefactor $1/(4\pi\epsilon_0)$ is omitted for simplicity.
The origin of the prefactor $1/2$ is explained in
Ref.~\cite{ballenegger2009simulations}.
The Ewald summation splits the Coulomb interaction into the direct, reciprocal and
correction contributions, i.e.~$E = E_\dir + E_\rec + E_\corr$ with
\begin{align}\label{eqn:es-dir}
  &E_{\dir}
   =
  \frac12 \sum^{\ast}_{\vect n}\sum_{i,j = 1}^{N}
  \frac{q_iq_j \erfc(\beta \vert\vect{r}_{ij} + \vect{n}\vert)}
  {\vert\vect{r}_{ij} + \vect{n}\vert},
  \\\label{eqn:es-rec}
  &E_{\rec}
   =
  \frac1{2\pi V} \sum_{\vect m \neq 0}
  \frac{\exp(-\pi^2\vect m^2 / \beta^2)}{\vect m^2}
  S(\vect m) S(-\vect m), \\\label{eqn:es-cor}
  &E_{\corr}
   =
  -\frac\beta{\sqrt \pi} \sum_{i=1}^N q_i^2.
\end{align}
We omit the surface energy term because the spherical summation order and
the metallic boundary condition are assumed to the system~\cite{ballenegger2009simulations,de1980simulation,de1980simulation2}.
In the Ewald summation, $\beta > 0$ is the splitting parameter that controls the
convergence speed of the direct and the reciprocal parts.
In the reciprocal energy~\eqref{eqn:es-rec},
$\vect m = m_1 \vect a_1^\ast + m_2 \vect a_2^\ast + m_3 \vect a_3^\ast$
with $(m_1, m_2, m_3) \in \mathbb Z^3$ and $(\vect a^\ast_1, \vect a^\ast_2, \vect a^\ast_3)$
being the reciprocal cell vectors defined by
$\vect a_\alpha \cdot \vect a_\gamma^\ast = \delta_{\alpha\gamma}$, where $\alpha,\gamma = 1,2,3$.
$V = (\vect a_1 \times \vect a_2) \cdot \vect a_3$ is the volume of the unit cell.
$
  S(\vect m) = \sum_{j=1}^N q_j e^{2\pi \imgunit \vect m\cdot \vect r_j}
$
is the structure factor, and the notation ``$\imgunit$'' at the exponent should be understood
as the imaginary unit (not a charge index). The magnitude of the structure factor is upper bounded by $\sum_j\vert q_j\vert$.

\newparagraph
In the direct energy part~\eqref{eqn:es-dir}, the complementary error function, i.e.~$\erfc$, converges exponentially
fast to zero with increasing charge distance $\vert \vect r_{ij}+\vect n \vert$, therefore,
it can be cut off and the direct energy is computed at the cost of $\mo(N)$
by using the standard cell division and neighbor list algorithms~\cite{frenkel2001understanding}.
The summand of the reciprocal energy~\eqref{eqn:es-rec} decays exponentially fast as $\vert \vect m\vert$ increases,
therefore, the infinite summation can be approximated by a finite summation, where $m_\alpha$
ranges from $-K_\alpha/2$ to $K_\alpha/2 - 1$ with $K_\alpha$ being the number of terms summed on direction $\alpha$.
A naive choice of $K_\alpha$ that preserves the accuracy  satisfies $K_1K_2K_3 \propto N$,
thus the computational complexity of the reciprocal energy is $\mo(N^2)$.

\newparagraph
The fast Ewald methods interpolate the point charges on a $K_1\times K_2\times K_3$ uniform  mesh,
then accelerate the computation of the structure factor, which is  a discretized Fourier transform of the charge distribution,
by the fast Fourier transform (FFT). 
Taking the SPME method as an example, the interpolation of the single particle charge contribution
to the mesh (on direction $\alpha$) reads~\cite{wang2016multiple}
\begin{align}\label{eqn:atom-intpl}
  q e^{2\pi \imgunit m_\alpha u_\alpha/K_\alpha}
  \approx
  \frac 1{K_\alpha\hmyphi(m_\alpha)}  \sum_{l\in I_K} q\, \myphi (u_\alpha - l) e^{2\pi \imgunit m_\alpha l / K_\alpha},
\end{align}
where $u$ is the scaled coordinate that is defined by $u_\alpha = K_\alpha r_\alpha$ with $r_\alpha = \vect a^\ast_\alpha \cdot \vect r$.
$I_K = \{ l\in \mathbb Z  : -K/2 \leq l < K/2\}$.
$\myphi$ is the interpolation basis that is usually assumed to be truncated with radius $C$,
which means the value of $\myphi$ out of the range $[-C, C]$ is assume to be 0.
The Fourier transform of $\myphi$ on $I_K$ is denoted by $\hmyphi$.
By using the approximation~\eqref{eqn:atom-intpl}, the reciprocal energy is proved to be~\cite{essmann1995spm} 
\begin{align}\label{eqn:spme-rec-1}
  E_\rec \approx
  \sum_{l_1,l_2,l_3}
  Q(l_1,l_2,l_3)
  [\,Q \ast (F B^2)^{\vee}] (l_1,l_2,l_3),
\end{align}
where ``$\ast$'' denotes the convolution, ``$\vee$'' denotes the inverse discrete
Fourier transform, and
\begin{align}\label{eqn:f}
  &F(\vect m)
  =
  \frac1{2\pi V}\times
    \begin{dcases}
      \frac{\exp(-\pi^2\vect m^2 / \beta^2)}{\vect m^2} &  \vert \vect m \vert \neq 0, \\ 
      \: 0 & \vert \vect m\vert  = 0,
    \end{dcases} \\\label{eqn:b}
  &B(\vect m) = \prod_\alpha \frac 1{ \hmyphi(m_\alpha)}, \\\label{eqn:p}
  &P_{\vect r}(l_1,l_2,l_3) = \prod_\alpha \myphi(u_\alpha - l_\alpha), \\ \label{eqn:q}
  &Q (l_1,l_2,l_3)= \sum_j q_j P_{\vect r_j} (l_1,l_2,l_3).
\end{align}
$Q(l_1,l_2,l_3)$ is the interpolated charge distribution on the mesh.
The interpolation of single particle to mesh, viz.~the computation of $P_{\vect r}(l_1, l_2, l_3)$, can be accomplished in $\mo(1)$ operations due to the
compact support of the interpolation basis $\myphi$,
thus the computational cost of $Q = \sum_j q_j P_{\vect r_j}$ is $\mo(N)$.
By using the identity
$Q \ast (F B^2)^{\vee} = [\, \hat Q \times (F B^2)\, ]^\vee$,
the computation of the convolution in Eq.~\eqref{eqn:spme-rec-1} is converted to a forward discrete Fourier transform of $Q$,
a multiplication between $\hat Q$ and $FB^2$,
and then a backward transform of $\hat Q \times (F B^2)$. 
The computational cost of the multiplication is $\mo(N)$ and that of the fast Fourier transforms is $\mo(N\log N)$,
thus the total computational cost of the reciprocal energy \eqref{eqn:spme-rec-1} is $\mo(N\log N)$.

\newparagraph
The reciprocal force of a charged particle can be computed in two ways.
The first way, known as \emph{ik-differentiation}, takes negative gradient of the reciprocal energy~\eqref{eqn:es-rec}
with respect to particle coordinate $\vect r$, then approximates the force by the particle-mesh interpolation. 
It leads to~\cite{darden1993pme}
\begin{align}\label{eqn:spme-ik}
  \vect F^\ik_{\rec,i}
  \approx \ &
              q_i
              \sum_{l_1,l_2,l_3}
              P_{\vect r_i} (l_1,l_2,l_3)
              [\,Q \ast (\vect G B^2)^{\vee}] (l_1,l_2,l_3),  
\end{align}
where
\begin{align}
  \vect G(\vect m) = -4\pi \imgunit\vect m F(\vect m).  
\end{align}
The second way, known as \emph{analytical differentiation}, takes negative gradient of the \emph{approximated}
reciprocal energy~\eqref{eqn:spme-rec-1}, and yields~\cite{essmann1995spm}
\begin{align}\label{eqn:spme-ad}
  \vect F^\ad_{\rec,i}
  \approx \ &
              q_i
              \sum_{l_1,l_2,l_3}
              - 2 \nabla_{\vect r_i} P_{\vect r_i} (l_1,l_2,l_3)
              [\,Q \ast (F B^2)^{\vee}] (l_1,l_2,l_3).
\end{align}

\newparagraph
In the energy approximation \eqref{eqn:spme-rec-1} and force approximations \eqref{eqn:spme-ik} and \eqref{eqn:spme-ad},
the only step that introduces errors is the particle-mesh interpolation~\eqref{eqn:atom-intpl}, 
thus the interpolation basis plays a crucial role in the accuracy of the fast Ewald methods.
{In the original work of the SPME and PPPM methods~\cite{essmann1995spm,hockney1988computer}, the B-spline basis was proposed for the interpolation.}
An $n$-th order B-spline basis is defined in a recursive way:
\begin{align}
  \myphi_n(x) = K \myphi_{n-1}\ast\myphi_1(x), \quad \myphi_1(x) = \chi_{[-1/2,1/2]}(x),
\end{align}
where $\chi_{[-1/2,1/2]}(x)$ is the characteristic function of interval $[-1/2,1/2]$.
The order $n$ B-spline basis has a compact support of $[-n/2, n/2]$, and is $(n-2)$-th order differentiable.
The basis truncation is usually taken as $C=n/2$.
The Kaiser-Bessel basis of truncation $C$ is defined by
\begin{align}
  \myphi (x) =
  \frac{\sinh (\pi \wshape \sqrt{C^2 - x^2} ) }{\pi\sqrt{C^2 - x^2}},
\end{align}
where $\wshape$ is the shape parameter, which can be determined by optimizing against
the reciprocal force error~\cite{gao2017kaiser}.

\section{The optimal interpolation basis}
\label{sec:opti}

The quality of the interpolation basis can be investigated by
evaluating the error introduced in the computation.
In the context of molecular dynamics (MD) simulations, the error is usually defined by the  root mean square (RMS) error of the reciprocal force computation, i.e.
\begin{align}\label{eqn:rms-error}
  \me = \sqrt{\langle \vert \Delta \vect F_\rec \vert^2\rangle}, \quad \Delta \vect F_\rec = \vect F_\rec - \vect F_\rec^\ast,
\end{align}
where  $\langle\cdot\rangle$ denotes the ensemble average, and 
$\vect F_\rec$ and $ \vect F_\rec^\ast$ denote the computed and exact reciprocal forces, respectively.
It has been shown that the RMS reciprocal error is composed of 
the homogeneity, the inhomogeneity and the correlation parts~\cite{wang2012numerical}
\begin{align}\label{eqn:esti-split}
  \vert \me\vert^2 =
  \vert \me_\homo \vert^2 + \vert \me_\inhomo \vert^2 + \me_\correlation.
\end{align}
The homogeneity error $ \me_\homo$ stems from the fluctuation of
the error force $\Delta \vect F_\rec$.
The inhomogeneity error $\me_\inhomo$ originates from the inhomogeneous charge distribution, 
and  has been shown to vanish when the system is locally neutral,
which is the case in most realistic systems. 
The correlation error $\me_\correlation$ contributes when the positions of
the charges are correlated.
For example, the partial charges in a classical point-charge water system
is correlated via the covalent bonds, the hydrogen bonds and the van der Waals interactions.

\newparagraph
An error estimate is an analytical expression of the RMS reciprocal force error
in terms of the working parameters including the splitting parameter, the mesh spacing and the basis truncation.
If the system is locally neutral and the positions of the charges are uncorrelated,
the inhomogeneity and the correlation errors vanish,
and the errors of the ik- and analytical differentiation force schemes are estimated by~\cite{wang2016multiple}
\begin{align}\label{eqn:esti-ik}
  \vert \me^\ik\vert^2 &=
  \vert \me^\ik_\homo\vert^2 \approx
  2 q^2 Q^2 \sum_{\vect m} \sum_{\alpha,l\neq 0} \mg^2_{\alpha,l}(\vect m),\\\label{eqn:esti-ad}
  \vert \me^\ad\vert^2 &=
  \vert \me^\ad_\homo\vert^2 \approx
  q^2 Q^2 \sum_{\vect m} \sum_{\alpha,l\neq 0} \mg^2_{\alpha,l}(\vect m) +
  q^2 Q^2 \sum_{\vect m} \sum_{\alpha,l\neq 0} (\mg_{\alpha,l} + \mf_{\alpha,l})^2(\vect m),
\end{align}
where $Q^2 = \sum_i q_i^2$, $q^2 = Q^2 / N$, and
we have the short-hand notations
\begin{align}\label{eqn:mg-def}
  \mg_{\alpha,l}(\vect m) &= \vect G (\vect m) Z_{\alpha,l}(\vect m),\\\label{eqn:mf-def}
  \mf_{\alpha,l}(\vect m) &= -4\pi \imgunit l K_\alpha \vect a_\alpha^\ast F(\vect m) Z_{\alpha,l}(\vect m), \\
  Z_{\alpha,l}(\vect m) &= {\hmyphi(m_\alpha + l K_\alpha)}/{\hmyphi(m_\alpha)}.
\end{align}

\newparagraph
The RMS reciprocal error estimates \eqref{eqn:esti-ik} and \eqref{eqn:esti-ad} clearly depend on the interpolation basis.
The optimal interpolation basis in terms of minimal numerical error 
is determined by solving the unconstrained optimization problem
\begin{align}\label{eqn:opti-orig}
  \min_{\myphi} \me [\myphi],
\end{align}
where $\me$ can be estimated by either \eqref{eqn:esti-ik} or \eqref{eqn:esti-ad} for ik- or analytical differentiation, respectively.
It is reasonable to assume that the interpolation basis is an even function on $[-C, C]$, thus only the value of $\myphi$ in range $[0, C]$ should be determined.
We uniformly discretize the range $[0, C]$ by $M$ nodes that are denoted by $x_i = i\mydx$ with $\mydx = C/M$,
and use the following ansatz to construct $\myphi$
\begin{align}\label{eqn:phi-discretize}
  \myphi(x) = \eta_0(x) + \sum_{i=1}^{M-1}\lambda_i \eta_i(x) + \sum_{i=1}^{M-1}\nu_i \theta_i(x),
\end{align}
where 
\begin{align}\label{eqn:def-eta-0}
  \eta_0 (x) &= H_{00}(\frac{x - x_{i}}{\mydx}), \\\label{eqn:def-eta}
  \eta_i (x) &= H_{01}(\frac{x - x_{i-1}}{\mydx}) + H_{00}(\frac{x - x_{i}}{\mydx}), \\\label{eqn:def-theta}
  \theta_{i} (x) &= \mydx H_{11}(\frac{x - x_{i-1}}{\mydx}) + \mydx H_{10}(\frac{x - x_{i}}{\mydx}).
\end{align}
$H_{00}$, $H_{01}$, $H_{10}$ and $H_{11}$ are cubic Hermite splines
(third order piecewise polynomials, details provided in Appendix~\ref{app:cubic-hermite}).
By this construction, the interpolation basis $\myphi(x)$ is first order differentiable and is normalized by $\myphi (0) = 1$.
The basis smoothly vanishes at the truncation $x = C$, viz.~$\myphi(C) = 0$ and $\myphi'(C) = 0$.
The prefactors $\{\lambda_i\}$ and $\{\nu_i\}$
are the values and derivatives of the basis at the discretization nodes,
because it can be proved that $\lambda_i = \myphi(x_i)$ and $\nu_i = \myphi'(x_i)$ (see Appendix~\ref{app:cubic-hermite} for details).
By inserting Eq.~\eqref{eqn:phi-discretize} into error estimates \eqref{eqn:esti-ik}
and \eqref{eqn:esti-ad}, the functionals of the interpolation basis $\myphi$ are converted into functions of $\{\lambda_i\}$ and $\{\nu_i\}$.
Thus the optimization problem~\eqref{eqn:opti-orig} is discretized as
\begin{align}\label{eqn:opti-disc}
  \min_{\{\lambda_i\}, \{\nu_i\}} \me(\{\lambda_i\}, \{\nu_i\}).
\end{align}
Since the error estimates \eqref{eqn:esti-ik-cont} and \eqref{eqn:esti-ad-cont} are positive definite and are  provided in squared form,
it is more convenient to solve the equivalent optimization problem
\begin{align}\label{eqn:opti-disc2}
  \min_{\{\lambda_i\}, \{\nu_i\}} \vert \me(\{\lambda_i\}, \{\nu_i\})\vert^2.
\end{align}
This is a classic unconstrained optimization problem, which can be solved by well
established algorithms such as conjugate gradient method or Broyden-Fletcher-Goldfarb-Shanno
(known as BFGS) method~\cite{fletcher1987practical}. Our code uses the implementation of the BFGS method from Dlib~\cite{king2009dlib}.

\section{The generality of the optimal basis}
\label{sec:generality}

The error estimates \eqref{eqn:esti-ik} and \eqref{eqn:esti-ad} 
depend on the system,
thus the optimization problem \eqref{eqn:opti-orig} should be solved for each specific system 
to obtain the optimal basis.
In this section, we demonstrate that there exists a general optimal basis 
that minimizes the estimated error in systems with different amounts of charge per particle,  different number of charges and different system sizes,
as long as the system is large enough.

\newparagraph
To begin with, it is more convenient to investigate the properties of the error estimates
if the summations in \eqref{eqn:esti-ik} and \eqref{eqn:esti-ad} are converted
to integrations, which are referred to as the continuous forms of the error estimates.
To simplify the discussion, we assume that the simulation region is cuboid.
We denote the mesh spacing on direction $\alpha$ by $h_\alpha =  L_\alpha / K_\alpha$, where $L_\alpha = \vert\vect a_\alpha\vert$,
and further assume that the mesh spacings are roughly the same on all directions,
i.e.~$h \approx h_1 \approx h_2 \approx h_3$.
The continuous form of the error estimate \eqref{eqn:esti-ik} is 
(see Appendix~\ref{app:cont} for details)
\begin{align} \label{eqn:esti-ik-cont}
  \vert \me^\ik_\homo\vert^2 \approx
  \frac{4 q^4\rho\beta}{(\beta h)^3}
  \int_{I^3}
  g^2\Big( {\vert \boldsymbol \mu\vert}/{(\beta h)} \Big)
  \sum_{\alpha,l\neq 0} 2z^2(\mu_\alpha, l)
  \:
  d\boldsymbol\mu,
\end{align}
where $\rho = N/V$ is the number density of charged particles in the system.
The integration region is defined by $I^3 = [-1/2,1/2] \times [-1/2,1/2] \times [-1/2,1/2]$.
We denote 
$\boldsymbol\mu = (\mu_1,\mu_2,\mu_3)$,
$g(m) = {e^{-\pi^2 m^2}} / {m}$,
and $z(\mu_\alpha, l) = {\tilde\myphi (\mu_\alpha+l)}/{\tilde\myphi(\mu_\alpha)}$ 
with $\tilde\myphi(\mu) = \int_{[-C,C]} \myphi(x)\ e^{-2\pi \imgunit \mu x} dx  $.
The continuous form of the error estimate \eqref{eqn:esti-ad} is
\begin{align} \label{eqn:esti-ad-cont}
  \vert \me^\ad_\homo\vert^2 \approx
  \frac{4q^4\rho \beta}{(\beta h)^3}
  \int_{I^3}
  g^2\Big( {\vert \boldsymbol \mu\vert}/{(\beta h)} \Big)
  \sum_{\alpha,l\neq 0}
  \Bigg[
  \Big(1 + \frac{\vert\boldsymbol\mu + l\vect e_\alpha\vert^2}{\vert\boldsymbol\mu\vert^2}\Big)
  z^2(\mu_\alpha, l)
  \Bigg]
  \:
  d\boldsymbol\mu.
\end{align}
It is noted that the continuous forms \eqref{eqn:esti-ik-cont} and \eqref{eqn:esti-ad-cont} do not depend on the system size.
As shown in Appendix~\ref{app:cont}, the standard error estimates \eqref{eqn:esti-ik} and \eqref{eqn:esti-ad}
are discretizations of  the continuous error estimates \eqref{eqn:esti-ik-cont} and \eqref{eqn:esti-ad-cont}, respectively,
with $K_\alpha$ being the number of discretization points on direction $\alpha$.
The difference between the standard and the continuous error estimates is
the error of the discretization, which vanishes as $K_\alpha$ goes to infinity.
It is noted that when taking the limit, the mesh spacing $h$ is fixed,
thus the limit implies that the system size $L_\alpha = h K_\alpha$ goes to infinity.
  Therefore, when the system is large enough, the optimal bases that minimize the standard error estimates
  converge to the bases that minimize the corresponding continuous error estimates, which are system-size independent.

\newparagraph
  Before further investigating the properties of the optimal basis, we will firstly show when a standard error estimate converges to its continuous form.
The standard error estimate
is close to the continuous form 
only when the variation of the integrand is resolved by enough discretization points.
In order to resolve the variation of $g^2({\vert \boldsymbol \mu\vert}/{(\beta h)})$, the characteristic size $\beta h / (\sqrt 2\,\pi)$ should be  discretized by enough points.
If two discretization points are required for the size $\beta h / (\sqrt 2\,\pi)$, 
then $K_\alpha \geq 2\sqrt 2 \pi / (\beta h)$, or
equivalently $L_\alpha = hK_\alpha \geq 2 \sqrt 2 \pi / \beta$.
This indicates that converged system size is inversely proportional to the splitting parameter $\beta$.
Taking the B-spline basis for instance, the convergence of the error estimate \eqref{eqn:esti-ik} with respect to the system size is
numerically investigated in Fig.~\ref{fig:esti-conv}.
When $\beta = 1.0\ \textrm{nm}^{-1}$, the system size should be larger than $4.5$~nm to have a converged error estimate.
When $\beta$ increases to $1.5\ \textrm{nm}^{-1}$, the minimal system size is $3.0$~nm.
The minimal system size reduces to 2.3~nm when $\beta = 2.0\ \textrm{nm}^{-1}$.
Although the minimal size is lower than the rough estimate $2\sqrt 2 \pi/\beta \approx 8.9/\beta$, 
the inversely proportional relation between the minimal size and the splitting parameter is confirmed.
It is also observed that the minimal system size does not depend on the basis truncation.

\begin{figure}
  \centering
  \includegraphics[width=0.45\textwidth] {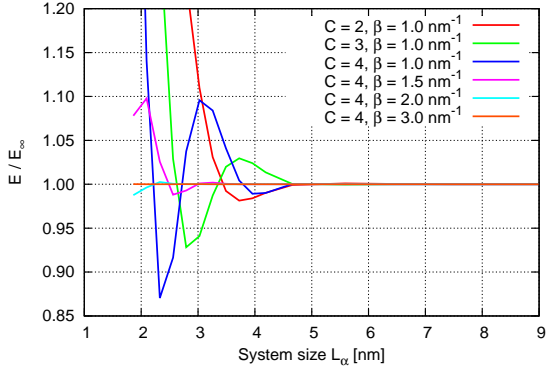}
  \caption{
    The convergence of the error estimate \eqref{eqn:esti-ik} to the continuous form \eqref{eqn:esti-ik-cont}.
    It is noted that the continuous form is system-size independent.
    The interpolation basis is B-spline. The mesh spacing is $h=0.117$~nm.
    We assume that $L_\alpha$ is the same on all directions,
    and plot the ratio between the error estimated in a system of size $L_\alpha$ and
    the error estimated in a fully converged system  of size $14.9$~nm.
    The lines with different colors denote different parameter choices as listed in the legend.
  } 
  \label{fig:esti-conv}
\end{figure}

\newparagraph
In this work, if not stated otherwise, we always assume that the system size is large enough,
so the error estimates \eqref{eqn:esti-ik} and \eqref{eqn:esti-ad} converge to the
continuous forms \eqref{eqn:esti-ik-cont} and \eqref{eqn:esti-ad-cont}, respectively,
and the optimal bases are also converged.
  The properties of an optimal basis can be analyzed
  by investigating the corresponding continuous error estimate.
For a given basis truncation $C$, the optimal basis has the following properties
\begin{enumerate} \setlength \itemsep{ 0pt }
\item The optimal basis is independent of the amount of charge per particle.
\item The optimal basis is independent of the number of charged  particles.
\item The optimal basis is independent of the size of the system.
\item The optimal basis is completely determined by the characteristic number $\beta h$.
\end{enumerate}
Property 1 holds
because the average amount of charge $q^2 = \sum_i q_i^2/N$ is a prefactor
of the error estimates \eqref{eqn:esti-ik-cont} and \eqref{eqn:esti-ad-cont}.
Property~3 holds because the error estimates are independent of the system size.
Property~2 holds because the number density $\rho$ is a prefactor of the error estimates, and because Property~3 holds.
When the truncation $C$ is fixed, the integrands of \eqref{eqn:esti-ik-cont} and \eqref{eqn:esti-ad-cont}
only depend on the number $\beta h$ via function $g^2({\vert \boldsymbol \mu\vert}/{(\beta h)} )$,
so the optimal basis is completely determined by this number.
Therefore, the optimal interpolation basis is system independent in the sense that
it applies to systems with different amounts of charge per particle, different numbers of charges and different system sizes.

\newparagraph
Due to the universality of the optimal interpolation basis,
the solutions to the optimization problem~\eqref{eqn:opti-disc2} are stored in a database.
The number of discretization nodes of the basis is set to $M = 40\,C$.
The mesh spacing is set to $0.117$~nm.
The number of mesh points is $K_\alpha = 64$ to ensure the convergence of error estimates.
The basis is optimized for a $\beta$
sequence starting from $1.0\ \textrm{nm}^{-1}$, increasing with a step of  $0.2\ \textrm{nm}^{-1}$, and ending at  $3.4\ \textrm{nm}^{-1}$,
and from $4.0\ \textrm{nm}^{-1}$, increasing with a step of  $1.0\ \textrm{nm}^{-1}$, and ending at  $7.0\ \textrm{nm}^{-1}$.
This provides the optimal bases for a $\beta h$ sequence that ranges from 0.117 to  0.818.
The optimal basis of the $\beta h$ that is not in the sequence is constructed by linear interpolation of neighboring optimal basis in the sequence.

\section{The numerical results}
\label{sec:result}

\begin{figure}
  \centering
  \includegraphics[width=0.45\textwidth] {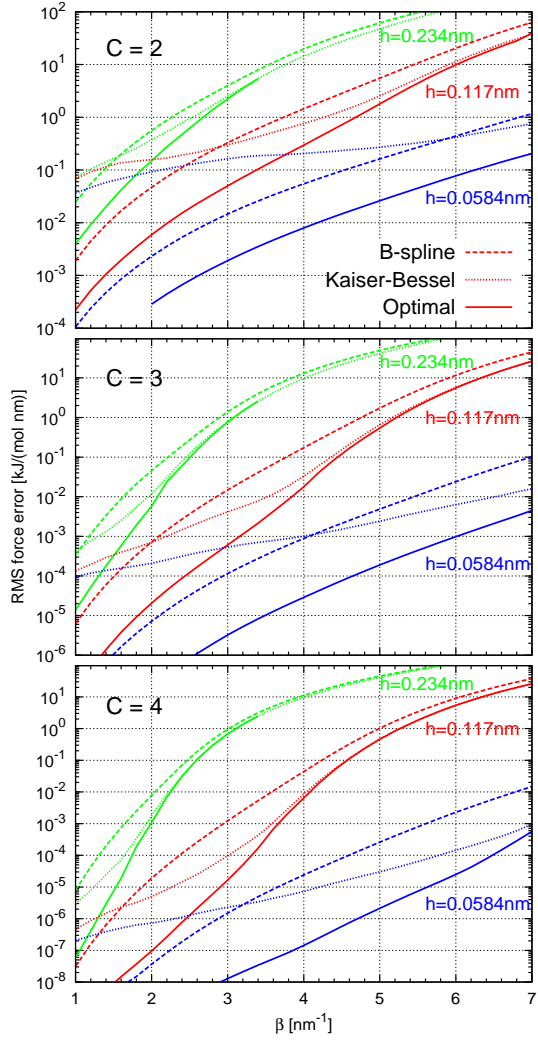}
  \caption[ik]{
    The RMS reciprocal force error
    of the B-spline (dashed lines), the Kaiser-Bessel (dotted lines) and the optimal (solid lines) bases
    plotted against the splitting parameter $\beta$ in the TIP3P water system.    
    The force scheme is the ik-differentiation.    
    From top to bottom, the three plots present the basis truncations of $C = 2$, 3 and 4, respectively.
    The green, red and blue lines present the mesh spacings of 0.234, 0.117 and 0.0584~nm, respectively.
  } 
  \label{fig:water-ik}
\end{figure}

\begin{figure}
  \centering
  \includegraphics[width=0.45\textwidth] {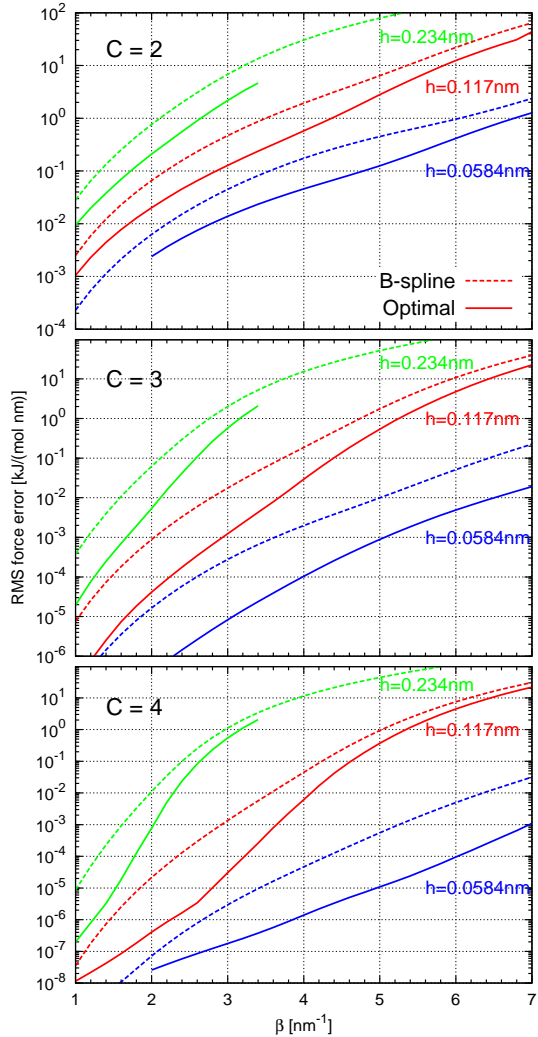}
  \caption[ad]{
    The RMS reciprocal force error
    of the B-spline (dashed lines) and the optimal (solid lines) bases
    plotted against the splitting parameter $\beta$ in the TIP3P water system.    
    The force scheme is the  analytical differentiation.    
    From top to bottom, the three plots present the basis truncations of $C = 2$, 3 and 4, respectively.
    The green, red and blue lines present the mesh spacings of 0.234, 0.117 and 0.0584~nm, respectively.
  } 
  \label{fig:water-ad}
\end{figure}

In this section, we investigate the RMS reciprocal force error of the
B-spline, the Kaiser-Bessel and the optimal bases in a TIP3P~\cite{jorgensen1983comparison} water system that has 13824 molecules.
Each water molecule is modeled by three point charges connected by  covalent bonds.
The oxygen atom has a partial charge of $-0.834\, e$
while the hydrogen atom has a partial charge of $0.417\,e$.
The O-H bond length is constrained to 0.09572~nm and the H-O-H angle is  constrained to $104.52^\circ$.
The simulation region is of size $7.48\,\textrm{nm} \times 7.48\,\textrm{nm} \times 7.48\,\textrm{nm}$, 
and is subjected to the periodic boundary condition.
The water configuration is taken from an equilibrated
NPT simulation~\cite{gao2016sampling}.
The computed reciprocal force is compared with a well converged Ewald summation,
and the RMS reciprocal force error is computed by definition~\eqref{eqn:rms-error}. 

\newparagraph
In Fig.~\ref{fig:water-ik}, we report the RMS force errors of the
B-spline (dashed line), the Kaiser-Bessel (dotted line) and the optimal (solid line) bases
using the ik-differentiation force scheme.
In Fig.~\ref{fig:water-ad}, we report the RMS force errors of the
B-spline (dashed line) and the optimal (solid line) bases
using the analytical differentiation force scheme.
In both figures, 
the error is plotted against the splitting parameter $\beta$.
The three plots, from top to bottom, present the results of basis truncations $C= 2$, 3 and 4, respectively.
In each plot, the green, red and blue lines  present the errors of the mesh spacings $h =  0.234$, 0.117 and 0.0584~nm, respectively.
In all cases, the optimal basis is
more accurate than the B-spline and the Kaiser-Bessel bases.
In some cases, the optimal basis achieves two orders of magnitude higher accuracy,
for example, the ik-differentiation with $C = 4$, $\beta = 3.0~\textrm{nm}^{-1}$ and $h = 0.0584$~nm.
The advantage of the optimal basis is observed to be more significant
for smaller splitting parameters, 
and
smaller mesh spacings.
Taking the ik-differentiation with $C=3$ and  $h =  0.117$~nm for example,
the optimal basis is 2.1, 9.7 and 35 times as accurate as the B-spline basis
and is 1.0, 1.9, and 34 times as accurate as the Kaiser-Bessel basis
at $\beta = 6.0$, 4.0 and $2.0~\textrm{nm}^{-1}$, respectively.
Taking $C = 3$ and $\beta = 3.0~\textrm{nm}^{-1}$ for example,
the optimal basis is 1.8, 24 and 35 times as accurate as the B-spline basis, 
and is 1.0, 6.9 and 165 times as accurate as the Kaiser-Bessel basis
at mesh spacings $h = 0.234$, 0.117 and 0.0584~nm, respectively.

\newparagraph
It is noted that the basis that was optimized at $h = 0.117$~nm is used for the simulations at
$h = 0.234$~nm and $h = 0.0584$~nm if the products
of $\beta h$ are the same.
For example, the basis optimized for $\{h = 0.117\,\textrm{nm},\ \beta = 2.0\,\textrm{nm}^{-1}\}$ is
used for the simulation at
$\{h = 0.234\,\textrm{nm},\ \beta = 1.0\,\textrm{nm}^{-1}\}$
and
at $\{h = 0.0584\,\textrm{nm},\ \beta = 4.0\,\textrm{nm}^{-1}\}$.
The bases were optimized for the $\beta h$ range from 0.117 to 0.818~nm,
thus, in Figs.~\ref{fig:water-ik} and \ref{fig:water-ad}, 
we do not have 
the optimal bases for $\beta > 3.5\,\textrm{nm}^{-1}$ at $h = 0.234$~nm,
nor the optimal bases for $\beta < 2.0\,\textrm{nm}^{-1}$ at $h = 0.0584$~nm.

\newparagraph
The number of floating point operations
attributed to the interpolation basis is
the number of floating point operations of each basis evaluation times 
the number of evaluations of the basis.
The numbers of evaluations are the same if two bases have the same truncation $C$.
The optimal basis is a $M = 40C$ pieces cubic piecewise polynomial, while
the B-spline basis is a $C$ pieces $(2C-1)$-th order piecewise poly-nominal.
For $C = 2$, each evaluation of the optimal basis needs as many floating point operations as the B-spline basis.
For $C \geq 3$ each evaluation of the optimal basis needs less floating point operations than the B-spline basis
\footnote{
The evaluation of Kaiser-Bessel basis requires a square root and a hyperbolic sine function, which are usually much more expensive than the polynomials.
It should be noted that, in the productive codes, the Kaiser-Bessel basis is usually implemented by cubic interpolation of tabulated values, 
thus each evaluation needs as many floating point operations as the optimal basis.
}.
On the other hand,
the number of the polynomial pieces of the optimal basis is larger than the B-spline basis,
thus the cache missing rate of the polynomial coefficients is likely to be higher than the B-spline basis.
We investigate the time-to-solution of the particle-mesh interpolation of the TIP3P water system on a desktop computer with
an Intel i7-3770 CPU and 32~GB memory. Only one core of the CPU was used in the tests.
The in-house MD software MOASP was compiled by GCC 4.7 with double precision floating point.
The force scheme was ik-differentiation, and the mesh spacing was 0.117~nm.
For the truncation $C=2$, the time-to-solutions of the optimal and the B-spline bases were
0.105 and 0.100~seconds, respectively.
The optimal basis was only 5\% slower than the B-spline basis.
For $C=4$, the time-to-solutions were
0.331 and 0.329~seconds, respectively.
In this case, the difference between the optimal and B-spline bases was less than 1\%.
In any case, the difference between the optimal and the B-spline bases in terms of time-to-solution
is not significant.

\section{Optimized with error estimate in correlated charge systems}
\label{sec:corr}

In Sec.~\ref{sec:opti},
the basis is optimized by minimizing the  estimated error that only includes the homogeneity error contribution.
This estimate, however, may not be able to precisely reflect the error in correlated charge systems.
Taking the TIP3P water system for example,
the covalently bonded atoms in one molecule have opposite charge signs and form a neutral charge group,
and the error is usually reduced by this charge correlation~\cite{wang2012numerical}.
The solution is to introduce the estimate of the correlation error 
to describe the reciprocal force error in a more precise way, i.e.
\begin{align}\label{eqn:esti-w-corr}
  \vert\me\vert^2 = \vert \me_\homo\vert^2 + \me_\correlation, 
\end{align}
It has been shown that, by using the TIP3P water system as an example,
introducing the bonded charge correlation in the error estimate leads to a substantial improvement of the quality of the estimate, 
and the improved estimate is good enough for the purpose of parameter tuning~\cite{wang2012numerical}.
The correlation errors of the ik- and analytical differentiations are estimated by~\cite{wang2016multiple}
\begin{align}\label{eqn:esti-corr-ik}
  \me^\ik_\correlation \approx &
  q^2 Q^2 \sum_{\vect m}\sum_{\alpha,l\neq 0}
  [T^w(\vert\vect m\vert) + T^w(\vert\vect m + l K_\alpha\vect a_\alpha^\ast\vert)] \mg^2_{\alpha,l}(\vect m), \\\nonumber
  \me^\ad_\correlation \approx &
  q^2 Q^2 \sum_{\vect m}\sum_{\alpha,l\neq 0}
  T^w(\vert\vect m\vert) (\mg_{\alpha,l} + \mf_{\alpha,l})^2(\vect m) \\\label{eqn:esti-corr-ad}
  &+
  q^2 Q^2 \sum_{\vect m}\sum_{\alpha,l\neq 0}
  T^w(\vert \vect m + l K_\alpha\vect a_\alpha^\ast \vert)\, \mg^2_{\alpha,l}(\vect m),
\end{align}
respectively, where  $\mg$ and $\mf$ are notations introduced by Eq.~\eqref{eqn:mg-def} and \eqref{eqn:mf-def}.
If the charge correlation due to the covalent bonds is considered, the term $T^w$ is defined by
\begin{align}\label{eqn:def-tw}
  T^w( m) 
  & =
    \frac{4\qh \qo}{2\qh^2 + \qo^2}
    \frac{\sin(2\pi m\so)}
    {2\pi m \so}
  +
    \frac{2\qh^2}{2\qh^2 + \qo^2}
    \frac{\sin(2\pi m \sh)}
    {2\pi m \sh}.
\end{align}
The notations $\qo$ and $\qh$ denote the partial charges of the oxygen and hydrogen atoms, respectively. 
{$\so$ denotes} the length of the covalent bond between the oxygen and hydrogen  atoms,
while {$\sh$ denotes} the distance between the two hydrogen atoms. 
Taking the TIP3P water model for example, $\qo = -0.834e$, $\qh =
0.471e$, $ \so  = 0.09572$~nm and $ \sh = 0.15139$~nm.

\newparagraph
The continuous form of the correlation error estimate for the ik-differentiation is given by
\begin{align}\label{eqn:esti-corr-ik-cont}
  \me^\ik_\correlation
  \approx
  \frac{4 q^4\rho\beta}{(\beta h)^3}
  \int_{I^3}
  g^2\Big( {\vert \boldsymbol \mu\vert}/{(\beta h)} \Big)
  \sum_{\alpha,l\neq 0}
  \Bigg\{
  \Big [
  T^w\Big(\frac{\vert \boldsymbol \mu\vert} h\Big) +
  T^w\Big(\frac{\vert \boldsymbol \mu + l\vect e_\alpha\vert} h\Big)
  \Big]
  z^2(\mu_\alpha, l)
  \Bigg\}
  \:
  d\boldsymbol\mu.
\end{align}
In the case of analytical differentiation, the continuous estimate is 
\begin{align} \nonumber
  \me^\ad_\correlation
  \approx &
            \frac{4q^4\rho \beta}{(\beta h)^3}
            \int_{I^3}
            g^2\Big( {\vert \boldsymbol \mu\vert}/{(\beta h)} \Big) \\\label{eqn:esti-corr-ad-cont}
          &
            \times \sum_{\alpha,l\neq 0} 
            \Bigg\{
            \Big[
            T^w\Big(\frac{\vert \boldsymbol \mu\vert} h\Big)
            \Big(1 + \frac{\vert\boldsymbol\mu + l\vect e_\alpha\vert^2}{\vert\boldsymbol\mu\vert^2}\Big)
            +
            T^w\Big(\frac{\vert \boldsymbol \mu + l\vect e_\alpha\vert} h\Big)
            \Big]            
            z^2(\mu_\alpha, l)
            \Bigg\}
            \:
            d\boldsymbol\mu.
\end{align}
If the system is large enough,
the standard estimates~\eqref{eqn:esti-corr-ik} and~\eqref{eqn:esti-corr-ad} converge to 
the continuous estimates, i.e.~\eqref{eqn:esti-corr-ik-cont} and~\eqref{eqn:esti-corr-ad-cont}, respectively.
Using similar arguments as those in Sec.~\ref{sec:generality},
the optimal basis taking into account the charge correlation has the properties as follows
\begin{enumerate} \setlength \itemsep{ 0pt }
\item The optimal basis is independent of the number of water molecules.
\item The optimal basis is independent of the system size.
\end{enumerate}
Unlike the optimal basis that only minimizes the homogeneity error,
the optimal basis considering the charge correlation is model specific,
because the value of the function $T^w$ in the integrands of~\eqref{eqn:esti-corr-ik-cont} and~\eqref{eqn:esti-corr-ad-cont}
depends on the amounts of the partial charge and the geometry of water molecule.
In the estimates \eqref{eqn:esti-corr-ik-cont} and \eqref{eqn:esti-corr-ad-cont},
the variation of the integrand is not only characterized by the dimensionless number $\beta h$,
but also by the dimensionless numbers  $\so / h$ and $\sh/h$ that indicate how fine the water geometry is resolved by the mesh.
Therefore, a basis optimized under a certain pair of the
splitting parameter $\beta$ and mesh spacing $h$ cannot be transferred to another pair with the same product.

\begin{figure}
  \centering
  \includegraphics[width=0.49\textwidth]{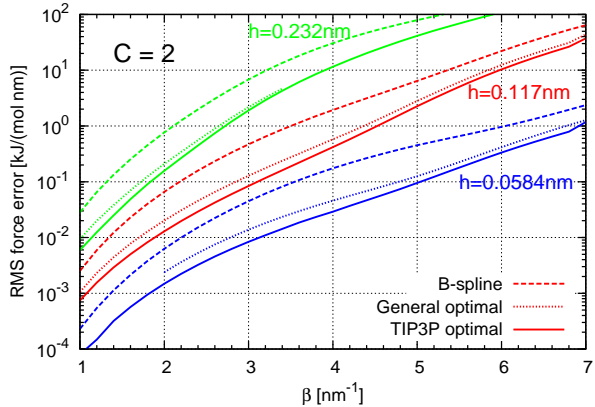}
  \caption{The RMS reciprocal force error of the B-spline (dashed lines),
    the general optimal (dotted lines) and the TIP3P optimal (solid lines) bases
    plotted against the splitting parameter in the TIP3P water system.
    The force scheme is the analytical differentiation, and the basis truncation is $C=2$.
    The green, red and blue lines present the mesh spacings of 0.234, 0.117 and 0.0584~nm, respectively.
  }
  \label{fig:err-corr}
\end{figure}

\newparagraph
Taking the TIP3P water system as an example,
we numerically compare the basis that optimizes the homogeneity error
and the basis that optimizes the estimated error including the charge correlation, i.e.~Eq.~\eqref{eqn:esti-w-corr}.
The former basis is model-independent, while the latter is model specific,
thus we refer to them as the general optimal basis and the TIP3P optimal basis, respectively.
In Fig.~\ref{fig:err-corr},
we present the accuracy of the B-spline basis (dashed lines), the general optimal basis (dotted lines)
and the TIP3P optimal basis (solid lines).
The basis truncation was set to $C=2$ for all cases.
The force scheme is analytical differentiation.
The green, red and blue lines represent the
errors of the mesh spacing $h = 0.234$, 0.117 and 0.0584~nm, respectively.
The general bases were optimized for different splitting parameter $\beta$ at $h = 0.117$~nm,
and were transferred to other mesh spacings if the products $\beta h$ are the same.
The TIP3P optimal bases were optimized for all the investigated combinations of the splitting parameter and mesh spacing.
It is observed that
the TIP3P optimal basis is more accurate compared with the general optimal basis, 
and the advantage is more obvious for a smaller mesh spacing.
Taking $\beta = 3.0\ \textrm{nm}^{-1}$ for example,
% the TIP3P optimal basis is 1.14, 1.53 and 1.65 times as accurate as the general optimal basis
the TIP3P optimal basis reduces the error by 13\%, 34\% and 39\% compared with the general optimal basis
at mesh spacings $h = 0.234$, 0.117 and 0.0584~nm, respectively.
It should be noted that
the cost of better accuracy is the model generality.
% the TIP3P optimal basis achieves better accuracy
% at the cost of generality, because this
The TIP3P optimal basis is specifically optimal
for the TIP3P water system or systems dominated by the TIP3P water.
It is not guaranteed that the TIP3P optimal basis
is also optimal for other water models or other molecular systems with different charge correlations.
In these systems, if the model specific optimal basis is not available, the general optimal basis is recommended.

\section{Conclusion}
\label{sec:conclusion}

In this manuscript, the optimal particle-mesh interpolation basis that minimizes the estimated RMS force error is
proposed for the fast Ewald method.
It is demonstrated that the optimal basis achieves significantly higher accuracy than the widely used B-spline basis for both the ik- and analytical differentiation force schemes,
at a cost of marginally (less than $5\%$) longer computational time.
We prove that the optimal basis is system independent, and is 
determined by a characteristic number that is the product of
the splitting parameter and the mesh spacing.  Therefore, it is
convenient to build a database of the general optimal bases and to integrate them
into existing MD packages.
By taking into account the charge correlation, the
accuracy of the optimal basis is further improved.  However, the cost of this
improvement is the generality.  We show that the optimal basis
derived in this way is specific to a molecular model, and should be
optimized for all possible combinations of the splitting parameter and
mesh spacing.
Therefore, the choice between the general optimal basis and the model specific optimal basis
is a trade-off between the generality and the accuracy.

\section*{Acknowledgment}
H.W. is supported by the National Science Foundation of China under Grants 11501039 and 91530322.
X.G. is supported by the National Science Foundation of China under Grant 91430218.
% The authors gratefully acknowledge the financial support from
% National High Technology Research and Development Program of China under Grant 2015AA01A304,
% and National Key Research and Development Program of China under Grant 2016YFB0201200.
The authors gratefully acknowledge the financial support from
the National Key Research and Development Program of China under Grants 2016YFB0201200 and 2016YFB0201203, 
and the Science Challenge Project No.~JCKY2016212A502.

\appendix
\section{The cubic Hermite splines }
\label{app:cubic-hermite}
The cubic Hermite splines $H_{00}$, $H_{01}$, $H_{10}$ and $H_{11}$ are defined,
on the interval $u \in [0, 1]$, by
\begin{align}
  H_{00}(u) & = (1+2u)(1-u)^2,  \\
  H_{01}(u) & = u^2 (3 - 2u), \\
  H_{10}(u) & = u (1-u)^2, \\
  H_{11}(u) & = u^2 (u-1).
\end{align}
It can be easily shown that
\begin{align}
  H_{00}(0) &= 1, \quad H_{00}(1) = H'_{00}(0) = H'_{00}(1) = 0, \\
  H_{01}(1) &= 1, \quad H_{01}(0) = H'_{01}(0) = H'_{01}(1) = 0, \\
  H'_{10}(0) &= 1, \quad H_{10}(0) = H_{10}(1) = H'_{10}(1) = 0, \\
  H'_{11}(1) &= 1, \quad H_{11}(0) = H_{11}(1) = H'_{11}(0) = 0.
\end{align}
Therefore, the ansatz functions $\eta_i(x)$ and $\theta_i(x)$
defined by Eq.~\eqref{eqn:def-eta-0}--\eqref{eqn:def-theta}
are supported on the interval $x\in[x_{i-1}, x_{i+1}]$, and 
have the following properties:
\begin{align}
  &\eta_i(x_{i}) = 1, \quad \eta_i(x_{i-1}) = \eta_i(x_{i+1}) = 0,\\
  &\eta'_i(x_{i-1}) = \eta'_i(x_{i}) = \eta'_i(x_{i+1}) = 0, \\
  &\theta_i(x_{i-1}) = \theta_i(x_{i}) = \theta_i(x_{i+1}) = 0, \\
  &\theta'_i(x_{i}) = 1, \quad \theta'_i(x_{i-1}) = \theta'_i(x_{i+1}) = 0.
\end{align}
Therefore, the interpolation basis given by Eq.~\eqref{eqn:phi-discretize}
has the properties:
\begin{align}
  &\myphi(0) =1, \quad \myphi(C) = 0, \\
  &\myphi'(0) = \myphi'(C) = 0, \\
  &\myphi(x_i) = \lambda_i, \\
  &\myphi'(x_i) = \nu_i. 
\end{align}

\section{Proof of the error estimate in the continuous form}
\label{app:cont}
For simplicity, we consider a simulation region of cuboid shape,
and denote $L_\alpha = \vert \vect a_\alpha\vert$, then
$\vect a_\alpha = L_\alpha \vect e_\alpha$, where $\vect e_\alpha$ is the unit vector
on direction $\alpha$.
In the reciprocal space, $\vect a_\alpha^\ast = (1/L_\alpha) \vect e_\alpha$.
The error estimate of the homogeneity error of the
ik-differentiation is
\begin{align}\label{eqn:esti-ik-sum}
  \vert \me^\ik_\homo\vert^2 =
  2 q^2 Q^2 \sum_{\vect m}
  \vect G^2(\vect m) \sum_{\alpha,l\neq 0}
  \Big\vert \frac{\hmyphi(m_\alpha + l K_\alpha)}{\hmyphi(m_\alpha)}\Big\vert^2.
\end{align}
The summation on the r.h.s.~can be considered as an approximation to the integration
\begin{align}\label{eqn:app-b2}
  \vert \me^\rec_\ik\vert^2 =
  2q^2 Q^2
  \int d\vect m
  \vect G^2(\vect m)
  \sum_{\alpha,l\neq 0}
  \Big\vert \frac{\hmyphi(m_\alpha + l K_\alpha)}{\hmyphi(m_\alpha)}\Big\vert^2.
\end{align}
We change the integration variable from $m_\alpha$ to $\mu_\alpha = m_\alpha / K_\alpha$, then
the error estimate \eqref{eqn:app-b2} becomes
\begin{align}\label{eqn:app-b3}
  \vert \me^\rec_\ik\vert^2 =
  2q^4
  \int d \boldsymbol \mu
  N K_1K_2K_3
  \vect G^2\Big( \sum_\alpha \mu_\alpha h^{-1} \vect e_\alpha \Big)
  \sum_{\alpha,l\neq 0}
  \Big\vert \frac{\tilde\myphi(\mu_\alpha + l)}{\tilde\myphi(\mu_\alpha)}\Big\vert^2,
\end{align}
where we used the identity $Q^2 = Nq^2$, and 
\begin{align}\label{eqn:app-b4}
  \vect m
  = \sum_\alpha m_\alpha \vect a_\alpha^\ast
  = \sum_\alpha \mu_\alpha K_\alpha \frac 1{L_\alpha} \vect e_\alpha
  = \sum_\alpha \mu_\alpha h^{-1}\vect e_\alpha.
\end{align}
In the last equation of \eqref{eqn:app-b4},
we assumed that the mesh spacing $h_\alpha = L_\alpha/K_\alpha$ is
roughly the same on all directions, i.e.~$h_\alpha \approx h$.
The notation $\tilde \myphi$ in Eq.~\eqref{eqn:app-b3} is
the Fourier transform of the interpolation basis represented by the new variable $\mu$:
\begin{align}\label{eqn:app-b5}
  \tilde \myphi(\mu) =
  \frac 1{K} \int_{-C}^{C} \myphi(x)\ e^{-2\pi i \mu x} dx
  =
  \frac 1{K} \int_{-K/2}^{K/2} \myphi(x)\ e^{-2\pi i m x/K} dx
  =
  \hmyphi(m).
\end{align}
In the second equation of \eqref{eqn:app-b5}, we noticed that the interpolation basis $\myphi$ is supported
on $[-C,C]$.
The function $\vect G$ is defined by
\begin{align}\label{eqn:app-b6}
  \vect G(\vect m) = - \frac{2 \imgunit \vect m}{ V} \frac{\exp(-\pi^2\vect m^2/\beta^2)}{\vect m^2}, \quad \vert \vect m\vert \neq 0.
\end{align}
It is easy to show that
\begin{align}\label{eqn:app-b7}
  NK_1K_2K_3 \vect G^2(\vect m)
  =  \frac{4\rho}{h^3} \frac{\exp(-2\pi^2\vect m^2/\beta^2)}{\vect m^2}, \quad \vert \vect m\vert \neq 0.
\end{align}
Inserting \eqref{eqn:app-b7} into the error estimate \eqref{eqn:app-b3} yields
\begin{align}\label{eqn:esti-ik-inte}
  \vert \me^\ik_\homo\vert^2 =
  \frac{8\rho q^4\beta}{h^3\beta^3}
  \int d\boldsymbol\mu
  \frac{\exp(-2\pi^2 \vert\boldsymbol\mu\vert^2 / (\beta h)^2)}{\vert\boldsymbol\mu\vert^2 /(\beta h)^2 }  
  \sum_{\alpha,l\neq 0}
  \Big\vert \frac{\tilde\myphi(\mu_\alpha + l)}{\tilde\myphi(\mu_\alpha)}\Big\vert^2.
\end{align}
The continuous error estimate \eqref{eqn:esti-ik-cont} is proved.
The continuous estimate of the homogeneity error of the analytical differentiation~\eqref{eqn:esti-ad-cont}, 
and the estimates of the correlation errors of the ik- and analytical differentiations~\eqref{eqn:esti-corr-ik-cont} and \eqref{eqn:esti-corr-ad-cont}
can be proved analogously.

\newparagraph
If we discretize the continuous form of the error estimate~\eqref{eqn:esti-ik-inte} by $K_\alpha$ discretization points on direction $\alpha$, 
we have, by replacing $d\mu_\alpha$ with $1/K_\alpha$ and $\mu_\alpha$ with $m_\alpha / K_\alpha$, 
\begin{align}\nonumber
  \vert \me^\ik_\homo\vert^2 =
  \frac{8\rho q^4\beta}{h^3\beta^3}
  \sum \frac1{K_1K_2K_3}
  \frac
  {\exp(-2\pi^2 \vert\sum (m_\alpha/K_\alpha)\vect e_\alpha\vert^2 /( \beta h)^2)}
  {\vert\sum (m_\alpha/K_\alpha)\vect e_\alpha\vert^2 /(\beta h)^2 }  
  \sum_{\alpha,l\neq 0}
  \Big\vert \frac{\hmyphi(m_\alpha + l K_\alpha)}{\hmyphi(m_\alpha)}\Big\vert^2,
\end{align}
where we have used Eq.~\eqref{eqn:app-b5}.
Noticing that $  \vect m = \sum m_\alpha/(hK_\alpha) \vect e_\alpha $ and the identity \eqref{eqn:app-b7}, 
the standard error estimate~\eqref{eqn:esti-ik-sum} is recovered.
Therefore, the standard error estimate is the discretization of the integration in the continuous form, 
and the difference will vanish as the number of discretization nodes $K_\alpha$ goes to infinity.

% \bibliography{ref}{}
% \bibliographystyle{unsrt}

\end{document}